\documentclass[conference]{IEEEtran}
\usepackage{mathrsfs}
\usepackage{amsmath}
\usepackage{amsthm}
\usepackage{tabularx}
\usepackage{booktabs}
\usepackage{amssymb}
\usepackage{graphicx}
\usepackage{times}
\usepackage{latexsym}
\usepackage{subfigure}
\usepackage{cite}
\usepackage{balance}
\usepackage{multirow}
\usepackage{makecell}
\usepackage{algorithmic}
\usepackage{color}
\usepackage{float}
\usepackage{cases}
\usepackage{algorithm}
\usepackage{yhmath}
\usepackage{footnote}

\usepackage[justification=centering]{caption}
\theoremstyle{remark}

\begin{document}
\title{An Overview of Low latency for Wireless Communications: an Evolutionary Perspective}
\author{Xin Fan, Yan Huo\\School of Electronics and Information Engineering, Beijing Jiaotong University, Beijing, China\\E-mail: \{fanxin, yhuo\}@bjtu.edu.cn}
\maketitle

\begin{abstract}
  Ultra-low latency supported by the fifth generation (5G) give impetus to the prosperity of many wireless network applications, such as autonomous driving, robotics, telepresence, virtual reality and so on. Ultra-low latency is not achieved in a moment, but requires long-term evolution of network structure and key enabling communication technologies. In this paper, we provide an evolutionary overview of low latency in mobile communication systems, including two different evolutionary perspectives: 1) network architecture; 2) physical layer air interface technologies. We firstly describe in detail the evolution of communication network architecture from the second generation (2G) to 5G, highlighting the key points reducing latency. Moreover, we review the evolution of key enabling technologies in the physical layer from 2G to 5G, which is also aimed at reducing latency. We also discussed the challenges and future research directions for low latency in network architecture and physical layer.
\end{abstract}
\begin{IEEEkeywords}
Low latency, physical layer, network architecture, evolution.
\end{IEEEkeywords}

\section{Introduction} 
With the development of mobile communication technologies, the requirements of human beings are increasing constantly. Once current requirements are met by communication technologies, new needs will arise and then new technologies will be expected to be updated. In this way, with the mutual promotion of requirements and communication technologies, we have been entering the fifth generation (5G) era. 5G is committed to creating a single platform to provide a wide range of services that are classified by the International Telecommunication Union (ITU) into three categories: enhanced mobile broadband (eMBB), massive machine-type communication (mMTC), ultra-reliable and low latency communications (URLLC) \cite{series2015imt}. To support these services, diverse sets of key performance indicators (KPIs) need to be achieved, which is a challenge for communication technologies. Among these KPIs, low latency that end-to-end (E2E) latency of 1 ms or less is perhaps the most challenging, due to the fact that most latency-sensitive services need to  simultaneously meet other KPIs, such as high reliability as $99.9999\%$ \cite{Parvez2018A}.

As mentioned above, the requirement for low latency is becoming more stringent step by step. The requirement for E2E latency of 1ms in 5G mobile systems is based on the fact that the fourth generation (4G) mobile communications can achieve $30-100$ ms E2E latency \cite{ji2018ultra}. The E$2$E latency of 4G is significantly improved on the basis of several hundred milliseconds of the third generation (3G) mobile communications that also developed from the previous second generation (2G). The current requirement for low latency cannot be satisfied by legacy mobile communication systems because of outdated inherent network architecture and communication techniques. In order to address requirement for the low latency, 5G needs to make a significant further evolution to the network architecture, while its relevant key enabling technologies require breakthrough innovation with respect to previous generations. Therefore, it is extremely important to trace back the evolution of the previous and current network architectures and communication technologies.
\subsection{Related Work}
In the existing literature, it can be easily found some overviews toward 5G networks, including network architecture \cite{habibi2019comprehensive, agiwal2016next, gupta2015survey}, physical layer technologies \cite{gupta2015survey, yuan20175g}. Apart from this, overviews on low latency for 5G networks in Internet \cite{Briscoe2016Reducing}, cloud computing \cite{srivastava2016survey}, Internet-of-Things (IoT) applications \cite{schulz2017latency}, and even a comprehensive survey of latency reduction solutions in cellular networks towards 5G \cite{Parvez2018A} can be also available. However, to the best of our knowledge, these overviews only cover 5G mobile communication systems, and there is no clear evolutionary route on how latency has being reduced step by step. In other words, horizontal overviews on low latency can be found, but longitudinal ones are missing.
\subsection{Contribution and Motivation}
In this paper, we provide an overview of the latency reduction for mobile communication systems from two different evolutionary perspective. Firstly, we discuss the reduction of latency by changing the network structure, including the radio access network (RAN), core network, and bearer network each generation network structure changes from 2G to 5G mobile communication system. Further, focusing on the physical layer, we present the communication technologies involved in each generation of mobile communication systems for attaining low latency, including packet size, frame structure, minimum transmission interval, modulation schemes, coding schemes and so on. Last but not the least, we also point out the major current challenges in reducing latency and possible future research directions in terms of network architecture and physical layer technologies.

Our motivation is to present a longitudinal and evolutionary perspective on the development of low-latency schemes, with a view to seeking further space through reflecting on history. It should be noted that although we give several evolutionary routes of network architecture and physical layer technologies, the detailed comparison between these solutions involved in each evolutionary route is not within our scope of this work.

The rest of this paper is organized as follows. Section II states the components of latency that includes the overall latency and physical layer latency. In the following Section III, we discuss the changes in network architecture from 2G to 5G, including radio access network, bearing network and core network. Then we present changes in the physical layer technologies for low latency in the Section IV. Following that, we point out the current challenges and future research directions of reducing latency in the Section V. Finally, this paper is summarized in the final Section VI. For convenience, a list of major abbreviations is presented in \textbf{Table~\ref{tab:abbreviations}}.

\begin{table}[hbtp]
  \centering
  \caption{List of the major abbreviations}
  \label{tab:abbreviations}
  \begin{tabular}{cl}
    \hline\\[-3mm]
    {\bf \small Abbreviation}&\qquad {\bf\small Definition}\\
    \hline
    RAN      & Radio access network\\
    E2E      & End-to-end\\
    GSM      & Global Systems for Mobile Communications\\
    GPRS     & General Packet Radio Service\\
    UMTS     & Universal Mobile Telecommunications System\\
    BTS      & Base transceiver station \\
    BSC      & Base station controller\\
    Cs       & Circuit switching\\
    Ps       & Packet switching\\
    RNC      & Radio network controller\\
    SGW      & Service gateway\\
    PGW      & Packet data gateway \\
    SGSN     & Service GPRS supported node\\
    GGSN     & Gateway GPRS supported node\\
    C-RAN    & Centralized, Cooperative, Cloud and Clean RAN\\
    BBU      & Base Band Unit \\
    RRU      & Remote Radio Unit\\
    NR       & New radio  \\
    SA       & Standalone  \\
    NSA      & Non-Standalone \\
    CU       & Centralized unit\\
    AAU      & Active antenna unit \\
    DU       & Distribute unit \\
    NFV      & Network function virtualization\\

    NFV      & Network function virtualization\\
    SDN      & Software defined network\\
    D2D      & Device-to-Device \\
    MSC      & Mobile switching center\\
    GMSC     & Gateway mobile switching center\\
    MGW      & Media Gateway\\
    MME      & Mobility management entity \\
    SBA      & Service Based architecture\\
    NF       & Network function  \\
    MEC      & Mobile edge computing  \\
    PDH      & Plesiochronous digital hierarchy \\
    SDH      & Synchronous digital hierarchy\\
    MSTP     & Multi-service transmission platform \\
    PTN      & Packet transport network   \\
    OTN      & Optical transport network \\
    PON      & Passive optical network\\
    NP       & Processor \\
    TM       & Traffic manager  \\
    TTI      & Transmission time interval  \\
    TDMA     & Time division multiple access  \\
    RLC      & Radio Link Control  \\
    EDGE     & Enhanced data rate for GSM evolution \\
    CDMA     & Code division multiple access\\
    LTE      & Long Term Evolution \\
    CP       & Cyclic prefix \\
    HSDPA    & High speed downlink packet access \\
    OFDM     & Orthogonal frequency division multiplexing  \\
    3GPP     & The 3rd Generation Partnership Project \\
    LDPC     & Low-density parity check \\
    IDMA     & Interleave division multiple access \\
    SCMA     & Sparse code multiple access   \\
    NOMA     & Non orthogonal multiple access  \\
    FBMC     & filter bank multi-carrier   \\
    UFMC     & Universal filtered multi-carrier  \\
    GFDM     & Generalized frequency division multiplexing  \\
    AUSF     & Authentication Server Function \\
    AMF      & Core Access and Mobility Management Function  \\
    SMF      & Session Management Function \\
    UPF      & User plane Function \\
    NEF      & Network Exposure Function     \\
    NRF      & NF Repository Function    \\
    PCF      & Policy Control function \\
    UDM      & Unified Data Management \\
    AF       & Application Function   \\
    NSSF     & Network Slice Selection Function \\

            \hline
  \end{tabular}
\end{table}

\section{Components of Latency}
It is important to understand the generation and composition of latency in order to better discuss the reduction of latency. In general, the cellar network latency can be divided into two aspects: 1) control plane latency; 2) user plane latency. The control plane latency generally refers to the time required for a terminal to switch from idle state to connected state; and the user plane latency refers to the time required for an IP message (ping packet) to be sent from a terminal to the application server and then returned to the terminal. Since the users' experience of network services mainly depends on the user plane latency (the control plane latency mainly affects network switching), low-latency communication is more focused on the user plane.

In terms of network architecture, the user plane latency consists of several components, including air interface, bearing network, core network and public data network (PDN)/Internet. As shown in Fig.1, the total unidirectional transmission latency can be expressed as:

\begin{equation}\label{T_total}
 T=T_{Radio}+T_{Bearing}+T_{Core}+T_{PDN}
\end{equation}
where
\begin{itemize}
  \item $T_{Radio}$ is the latency from the user terminal to the radio access network. This part of the latency is also known as the air interface latency, which is mainly affected by the physical layer transmission.
  \item $T_{Bearing}$ is the latency for transmission on the bearing/backhaul network that bears the connection between the radio access network and the core network, and between the core network and the PDN.
  \item $T_{Core}$ is the processing latency inside the core network. The processing includes mobility management, users' IP address allocation, security management, bearer control, etc.
  \item $T_{PDN}$ is the latency of content delivery for PDN to process requests and establish default bearers.
\end{itemize}

Obviously, the E$2$E latency is approximately twice as long as the above latency, i.e., $2\times T$.
As the main part of $T_{Radio}$, the physical layer transmission latency can be divided into five distinct components as follows:
\begin{equation}\label{T_pl}
 T_{PL}=T_{que}+T_{ttt}+T_{proc}+T_{prop}+T_{retr}
\end{equation}
where
\begin{itemize}
  \item $T_{que}$ is the queuing latency that is the time needed for the current packet to wait for the completion of the transmission of the previous packet. The queuing latency of a particular packet depends on the number of packets arriving in advance and waiting for transmission to the link. If the queue is empty and no other data packets are currently being transmitted, the queuing latency of the data packet is 0. 
  \item $T_{ttt}$ is the time-to-transmission latency that is the time required to push all the bits of a data packet to the link (from the first bit of the transmitted data packet to the last bit of the packet).
  \item $T_{proc}$ is the processing latency including encoding and decoding, modulation and demodulation, channel interleaving, channel estimation, rate matching, layer mapper, scrambling, data and control multiplexing, etc. These depend not only on physical layer technologies, but also on the processing capacity of user terminals and base stations.
  \item $T_{prop}$ is the propagation latency that is the time it takes for electromagnetic wave to propagate a certain distance in the channel.
  \item $T_{retr}$ is the latency of retransmission. Low link reliability can easily result in packet loss, which involves retransmitting.
\end{itemize}

E2E latency is the sum of latency on multi-segment paths. It can not satisfy the extreme latency requirement of 1 ms only by optimizing a local latency. Therefore, the implementation of 5G ultra-low latency requires a series of organically combined technologies. On the one hand, evolutionary changes in network architecture are needed to flatten the network structure and sink content providers. On the other hand, it is necessary for air interface be reconstructed to greatly reduce the transmission latency of physical layer. The vision of reducing latency can not be achieved overnight, but requires long-term efforts.

In the following two sections, we will discuss how to reduce latency from two aspects of the network architecture and physical layer air interface technologies in an evolutionary perspective, respectively.
\section{Evolutionary Network Architecture for Low Latency}\label{sec:NetArch}
\begin{figure*}[htbp]
\centering
\includegraphics[scale=0.45]{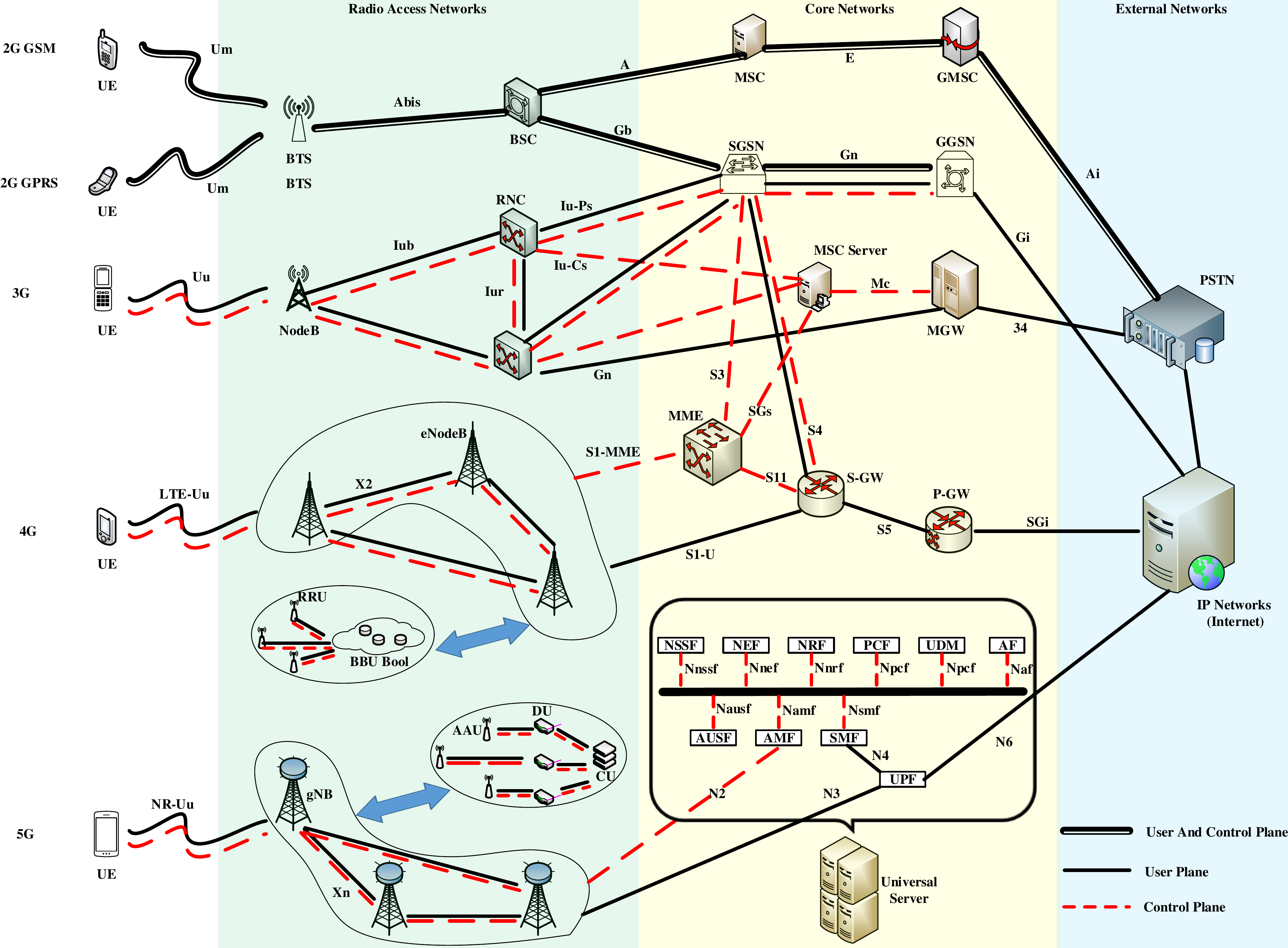}
\caption{The network architecture from 2G to 5G}
\label{fig_network}
\end{figure*}

Mobile communication networks have different architectures in different periods. Every generation of network architecture change is an innovation or evolution, which may be very wide-ranging for different requirements. This paper only highlights the significant changes in reducing latency, which may not necessarily provide a comprehensive overview of changes in network architecture. In this section, we provide an evolution of network architecture from 2G to 5G \footnote{The first generation mobile communication (1G) only provides voice service with analog signals, but no data transmission. Therefore, this paper ignores the discussion of 1G network architecture.}, including three parts: radio access network, core network and bearing network, as shown in Fig. \ref{fig_network}.

\subsection{Radio Access Network}
In fact, the earliest 2G network is known as Global Systems for Mobile Communications (GSM) networks, which also does not support data transmission, but only used digital signals to provide telephone services. From GSM network to General Packet Radio Service (GPRS) networks (called as 2.5G), packet switching service was introduced, and then began to provide data service. Therefore, it should be noted that the evolution of network architecture in terms of latency reduction starts at 2.5G.

\subsubsection{2G Radio Access Network}
The radio access network in 2G networks is composed of Base Transceiver Station (BTS) and Base Station Controller (BSC). BTS receives the wireless signal from mobile station (MS) through the Um air interface, then transmits it to BSC through the Abis interface. BSC is responsible for the management and configuration of wireless resources (such as power control, channel allocation, etc.), and then transmits the received signal to the core network through the A interface.
\subsubsection{2G-2.5G Radio Access Network Evolution}
The original GSM network is based on circuit switching technology and does not have the function of supporting packet switching services. Therefore, in order to support packet services, several functional entities have been added to the original GSM network structure, which is equivalent to adding a small network on the basis of the original network to form a GPRS network. For the radio access network, Packet Control Unit (PCU) is added to the BSC to provide packet switching channel. Starting from the GPRS network structure, two concepts are introduced, as follows. 
\begin{itemize}
  \item Circuit switching (Cs) domain: based on circuit switching, mainly including voice services, also including circuit-based data services, the most common is the fax service;
  \item Packet switching (Ps) domain: based on packet switching, mainly for common data services, including streaming media services, voice over IP (VOIP) and so on.
\end{itemize}

\subsubsection{2.5G-3G Radio Access Network Evolution}
 The 3G mobile cellular system has several standard, among which the Universal Mobile Telecommunications System (UMTS) is currently the most widely used. UMTS, sometimes also referred to as 3GSM, emphasizes the integration of 3G technologies and is a successor to the GSM standard. And packet switching system in UMTS is evolved from GPRS system, so the architecture of the system is quite similar but not exactly the same.

Instead of BTS and BSC, the composition of the radio access network is replaced by NodeB and Radio Network Controller (RNC). The functions of BTS and NodeB are basically the same, but there are still some differences between them. The differences are instrumental in reducing the latency as follows:
\begin{itemize}
  \item The interface between BTS and BSC is Abis, while the interface between NodeB and RNC is Iub. In physical layer transmission, Abis is a private interface that supports either ATM (E1/T1) or IP, while Iub is a open interface that supports ATM, IP, and Hybrid (ATM and IP). Because these interfaces are logical, many interfaces can be multiplexed and merged into a single transmission line. This reduces the number of interfaces and facilitates the switching efficiency between lines, which is conducive to the reduction of latency.
  \item In GSM, the overwhelming majority of control functions are implemented by BSC. BTS only plays the function of completing the coding and transmission of physical layer according to the instructions of RNC, and BTS itself basically has no ability to control and schedule physical resources. However, in order to achieve greater throughput capability on the air interface, the functions of NodeB are enhanced. The concepts of physical layer retransmission, spreading/despreading, fast resource scheduling and the closed (inner) loop power control, are introduced at the NodeB level. By introducing these functions on NodeB closer to the air interface, which should be only available with RNC, the efficiency of retransmission and air resource scheduling is accelerated. As a result, by sinking several control functions from RNC to NodeB, lower latency is achieved.
\end{itemize}

In addition to the changes in NodeB, there are also improvements between RNC and BSC that are beneficial to reducing the latency, as follows
\begin{itemize}
  \item The PCU entity was removed and its functions were incorporated into RNC, which helps to reduce latency.
  \item The interfaces between the radio access network and the core network are A interface in GSM and Gb interface in GPRS, respectively, whereas the two interfaces are consolidated into the Iu interface in UMTS. The Iu interface connecting the RNC to circuit-switched core network is known as Iu-CS and the Iu interface connecting the RNC to packet-switched core network is known as Iu-PS. Iu-CS and Iu-PS interfaces share the control plane using Radio Access Network Application Part (RANAP) protocol and have similar user planes \cite{muller2001further}. Therefore, the conversion between interfaces is easier and the latency is reduced.
  \item In GPRS, there is no connection between BSCs, whereas RNCs can be interconnected through the newly added Iur interface in UMTS. Benefiting from Iur interfaces, RNCs have the ability of macro diversity and can provide soft handover. Meanwhile, interaction efficiency between entities in access network has been improved, so as to reduce the latency for resource scheduling.
  \item In GPRS, some radio-related management functions are controlled in the core network, whereas these functions have been moved from the core network into the radio access network in UMTS \cite{Lin2010Mobility}. This change makes the radio access network and the core network more independent in function, and reduces the latency of signaling interaction. Due to the closer distance between the radio management functional entity and user terminals, the latency is also lower.
\end{itemize}

Apart from above changes, the Um air interface in the previous network structure was replaced by Uu interface. However, without air interface protocol and technology, the change of interface is of little significance. Thus, we will describe the evolution of physical layer air interface technologies in the next section. Here, we emphasize that Iub, Iur and Iu (including Iu-CS and Iu-PS) interfaces realize the separation of control plane and user plane in the protocol. Control plane latency is different from user plane latency (the number of control signaling is related to the number of users, while the amount of user data is largely related to new services and applications, as well as the performance of devices. The data quantity is different, the latency is different naturally). Thus, separating them and optimizing them separately is more beneficial to reduce network latency.

\subsubsection{3G-4G Radio Access Network Evolution}
The evolution of the architecture of 3G-4G radio access network can be summarized as follows:
\begin{itemize}
  \item One layer less: The RNC is removed from the 4G radio access network, which means that the whole network structure is reduced by one layer, namely, flattening. A more flat network structure reduces the complexity of the system, and reduces the multi-node overhead of information interaction between base stations and core networks. Therefore, flattening is beneficial to reduce latency. Not only the latency of user plane is greatly shortened, but also the time of state migration is reduced, due to the fact that the process of control plane from sleep state to activation state is simplified. 
  \item One interface more: 4G radio access network consists of several Evovled NodeBs (eNodeBs), with an X2 interface added between eNodeBs. In 3G, there were only interfaces (i.e., Iur) between RNCs, while no interfaces between NodeBs. Once the connection between NodeB and RNC is interrupted, the NodeB will become a lone point. However, because one base station is connected to many base stations, any fault between two points can also be connected through other channels, which will not cause a base station to become a lone point. The interface X2 between eNodeBs facilitates 4G wireless network to coordinate the work between the network elements and enhances the robustness of networks. This efficient and robust network cooperation is conducive to the reduction of latency.
  \item ``Fat" base station: In UMTS, NodeB is responsible for radio frequency processing and baseband processing. RNC is mainly responsible for controlling and coordinating the cooperation between base stations, including system access control, bearer control, mobility management, wireless resource management and other control functions. After removing RNC from eUTRAN, its underlying functions are allocated to eNodeB, and its high-level functions are allocated to AGW (access gateway, including service gateway SGW and packet data gateway PGW) of core network. The functions of eNodeB are mainly evolved from the functions of NodeB, RNC, the service GPRS supported node (SGSN) and the gateway GPRS supported node (GGSN) of 3G. That is to say, Some functions of the core network sink into the access network again, which results in lower latency.
  \item Centralized, Cooperative, Cloud and Clean radio access network (C-RAN): An eNodeB consists mainly of a Base Band Unit (BBU) and a Remote Radio Unit (RRU). A BBU can support multiple RRUs. In C-RAN, RRUs extend to the areas closer to user terminals through optical fiber, while BBUs are centralized to form BBU resource pools. By this, C-RAN centralizes all or part of the baseband processing resources, and uniformly manages and dynamically distributes them. While improving resource utilization and reducing energy consumption, C-RAN improves network performance through effective support for collaborative technologies. Due to the faster resource scheduling capability provided by C-RAN, the processing latency can be reduced.
\end{itemize}
\subsubsection{4G-5G Radio Access Network Evolution}
When 5G network is deployed in the existing network, it can be divided into two forms:
 \begin{itemize}
   \item Standalone (SA): refers to the reconstruction of 5G networks, including new base stations, backhaul links and core networks. While introducing new network elements and interfaces, SA will also adopt new technologies such as network virtualization and software-defined networks on a large scale, and integrate them with 5G new radio (NR). At the same time, the technical challenges faced by SA in protocol development, network planning, deployment and interoperability will surpass those faced by 3G and 4G systems.
   \item Non-Standalone (NSA): refers to the deployment of 5G networks using existing 4G infrastructure. The 5G carrier based on NSA architecture only carries user data, and its control signal is still transmitted through the 4G network.
 \end{itemize}

In this paper, we only consider the case of SA.
In 5G network, C-RAN still maintains the characteristics of the four ``C" but also has some significant evolution for low latency.

Firstly, because the demands in 5G is diversified, the network needs to be diversified; because the network needs to be diversified, it needs to be sliced; because it needs to be sliced, the network elements need to be able to move flexibly; because the network elements need to move flexibly, the connections between the network elements also needs to be flexible. As a result, 5G radio access network is redefined as NG-RAN, in which the base station is no longer eNodeB but gNB. The gNB reconstructs BBUs and RRUs into the following three functional entities:
\begin{itemize}
  \item Centralized Unit (CU): The non-real-time part of the original BBU will be separated and redefined as CU, which is responsible for handling non-real-time protocols and services.
  \item Active Antenna Unit (AAU): Part of the physical layer processing functions of original BBU is combined with the original RRU and passive antenna to form AAU.
  \item Distribute Unit (DU): The remaining functions of the original BBU are redefined as DU, which handles physical layer protocols and real-time services.
\end{itemize}

According to different service requirements and performance indicators, the network is divided into logical combinations of network functional entities, and the sliced network is used to provide specified services for target users and terminals. The reconstructed three functional entities coincide with the implementation of slicing. According to the 5G standard, CU, DU and AAU can be separated or co-located, so there will be a variety of network deployment patterns. In low-latency scenarios, DU needs to be deployed close to user terminals. The deployment of units that handle real-time services independently close to user terminals can greatly reduce latency.

Secondly, Network Function Virtualization (NFV) and Software Defined Network (SDN) are introduced into 5G networks. On the one hand, with the introduction of NFV, 5G network is constructed into a virtualized network environment. After virtualization, differentiated software functions run on the same hardware devices, and different network functions will share hardware computing, storage and communication resources. On the other hand, the introduction of SDN improves the network's programmability and separates the data and control aspects of the network. Under the NFV/SDN network architecture, resource pools that are inherited from 4G can be evolved into virtualized cloud resource pools (VCRPs). Resource allocation in VCPR can maximize the reuse of network resources, and bring more flexible and rapid sharing capability. NFV/SDN packages a series of network functions into a single action to minimize network sessions, which means a reduction in latency.

Thirdly, compared with cloud computing nodes, fog computing nodes are closer to user terminals. Therefore, introducing fog computing into wireless access network to build fog RAN (F-RAN) can alleviate the pressure of forward/backhaul links by virtue of the computing and caching potential of users and edge devices. In such a F-RAN network architecture, the distributed computing and caching capabilities of network edges can be effectively integrated through collaboration, which enhances the local real-time processing, transmission and control capabilities. By carrying sinking network functions and edge applications, local information processing and service distribution can be realized, providing lower the E2E latency performance.

Last but not the least, Device-to-Device (D2D) communication is not a new technology proposed by 5G, but D2D is destined to develop in 5G. Network participants share part of their hardware resources, including information processing, storage and network connectivity. These shared resources provide services and resources to the network and can be accessed directly by other users without passing through intermediary entities. This kind of direct communication mode will greatly reduce the communication E2E latency.
\subsection{Core Network}
The core network is the ``management center", which is mainly responsible for managing data, sorting data, and then distributing data. The functions implemented by each generation of core network are slightly different, which also corresponds to the changes of architecture. We enumerate the structural changes for latency reduction, mainly involving the sinking of functions and the separation of control and user plane.
\subsubsection{2G-2.5G Core Network}
In GSM, the core network is mainly composed of Mobile Switching Center (MSC), Visit Location Register (VLR), Home Location Register (HLR), Authentication Center (AUC), Equipment Identity Register (EIR) and other functional entities. MSC is the core, responsible for dealing with the specific service of users. VLR and HLR are mainly responsible for mobility management and user database management functions. AUC and EIR are responsible for security functions. In addition, the Gateway Mobile Switching Center (GMSC) is responsible for providing access to external network interfaces.

In the core network of GPRS, SGSN and GGSN are added, whose functions are consistent with MSC and GMSC, except that they deal with packet services and external network access to IP network, respectively.
\subsubsection{2.5G-3G Core Network}
The most significant change of 3G core network is the introduction of softswitch to separate the call control function from the media gateway (transport layer) in CS domain. The basic call control function is implemented by software, which realizes the separation of call transmission and call control, and establishes a separate plane for control, switching and software programmable functions. Specifically, the bearer and control functions of the MSC are separated and divided into two nodes, MSC-server and Media Gateway (MGW). Call control, mobility management, and media control functions are performed on the MSC-server, while the service bearer and media conversion functions are completed on the MGW. The biggest change brought by the structure of the separation of bearer and control is that MSC-servers and MGWs can be deployed separately. MSC-servers are usually concentrated in provincial capitals or regional centers. The centralized management of MSC-servers can improve the efficiency of operation and maintenance, while the MGW can be set according to the best service point. This centralized management and distributed service delivery network architecture is conducive to providing better services (including low latency).

Compared with CS domain, the PS domain only separates the user plane from the control plane logically, but not physically. The 3G core network in PS domain does not have separate entities to implement control plane and user plane, respectively \footnote{Direct tunnel technology is an innovative technology in 3G \cite{shaheen2007method,2009The}, which this paper does not focus on because of its uniqueness. The direct tunnel technology refers to the establishment of a ``direct channel" from RNC to GGSN. User plane data is transmitted in the ``direct channel" without passing through SGSN to realize the flattening of network plane.}.

\subsubsection{3G-4G Core Network}
The evolution of 4G core network has two main aspects as follows:
\begin{itemize}
  \item The CS domain is removed and the network architecture of a single PS domain is implemented. This single network architecture reduces signaling interaction and thus reduces latency.
  \item The control plane and the user plane are completely separated, physically and logically. The functions of control plane and user plane are assumed by different network entities respectively. The control plane element is Mobility Management Entity (MME), which is mainly used for user access control and mobility management. The user plane network element is the System Architecture Evolution-Gateway (SAE-GW), including Service-Gateway (S-GW) and Packet Data Network-Gateway (P-GW), is mainly used to bear data services. The processing efficiency of the core network is improved, so the latency is reduced.
  \end{itemize}

\subsubsection{4G-5G Core Network}
With the complexity and diversity of services in 5G era, the integrated network element structure of 4G core network can not flexibly cope with the changing service applications. In order to make the network elements more flexible and better respond to diversified applications, the 5G core network is evolving to a discrete Service Based architecture (SBA), which has two characteristics:
 \begin{itemize}
   \item Firstly, the separation of network functions absorbs the original design idea of NFV cloud, hoping to build the network in a way of software-based, modularized and service-oriented.
   \item Secondly, the user-side functions are free from the ``centralization" constraints, so that they can not only be flexibly deployed in the core network, but also can be deployed in the radio access network.
    \end{itemize}

Both of the above characteristics are beneficial to the reduction of latency. In order to achieve these two characteristics, the 5G core network has the following two evolutions from 4G:
\begin{itemize}
  \item Traditional network entities are split into multiple network functions (NFs) modules. In line with the concept of SBA, each NF is independently autonomous, and individual changes do not affect other NFs. The functions of the 4G core network elements can be found in the NFs of the 5G core network, but the architecture has changed from monolithic to micro-service. The most obvious external manifestation of this change is the substantial increase in network elements. These elements seem a lot, in fact, the hardware is virtualized in the virtualization platform. The purpose of this change is to make the network more flexible, open and scalable, so as to realize network slicing and provide better services.
  \item The traditional point-to-point communication between network elements is abandoned. The interface of each NF is a service interface. Each NF provides services through its own service-oriented interface, and allows other authorized NFs to access or invoke their own services. Because the underlying transport protocols are the same, all service interfaces can be transmitted on the same bus, that is, bus communication mode. This bus communication mode can provide higher information transmission efficiency and lower latency.
\end{itemize}

As the network elements are subdivided, the network elements on the user plane can further sink to mobile edge computing (MEC) nodes. The MEC technology enables applications, services and content to be deployed locally, near-by and distributed by migrating computing storage and service capabilities of the core network to the edge of the network. Benefiting from this, the content of service caching is close to the user terminal device, thus greatly reducing the service connection and response latency \footnote{After introducing MEC technology, by superimposing MEC servers on the base station side, content extraction and caching can be accomplished directly by MEC servers. In this way, when other terminals within the same base station call the same content, they can obtain directly from MEC servers. No more duplicate acquisition through the core network, which effectively saves the system resources on the core network side. At the same time, due to the sinking of service content, the corresponding service response latency will be significantly shortened.}.

\subsection{Bearer Network}
Bearer networks, sometimes referred to as transport networks or backhaul networks, are responsible for bearing and transmitting information. Generally speaking, the bearer network is the connecting part between the access network and the core network. In fact, the connections between the internal nodes of the radio access network and the core network should also be included. Although the bearer network is not the main target of low latency improvement, it also has to undertake some improvements in low latency. The latency of bearer network comes from two parts: 1) the time of signal propagation in medium; 2) the forwarding latency of transmission devices. We list some macro-evolutionary measures for lower latency as follows.
\subsubsection{Transmission distance}
As mentioned above, the functions of core networks are sinking from 2G to 5G networks. What's more, the application of MEC enables some low-latency services to be implemented directly in MEC devices without going through the core network. In this way, the transmission distance of the bearer network will be reduced, which means the transmission latency will be reduced. In addition, the flattening of the network architecture reduces the number of entities and forwarding hops, thus reducing the forwarding latency.
\subsubsection{Transmission media}
 In the early stage of communication development, T1/E1 copper lines (circuit switching) was used in bearer networks. With the rapid increase of mobile devices, the development of 3G technologies have brought tremendous operational expenditure (OPEX) pressure to bearer networks. Due to its low price and some other advantages (larger transmission bandwidth, larger channel capacity, lower line loss, longer transmission distance, stronger anti-interference ability, etc.), optical fiber transmission has been widely used. In the 4G network, removing the CS domain to promote ``All-IP", optical fiber has become the main force of bearer networks. However, there are still some electrical nodes in 4G networks, which cause some performance (including latency) losses due to the photoelectric conversion between optical nodes and electrical nodes. Thus, 5G puts forward the concept of all-optical network (referring to the electrical/optical and optical/electrical conversion of signal only when it comes in and out of the network) to improve network performance.

 On the other hand, with the rise of millimeter wave and large-scale Multiple-Input Multiple-Output (MIMO) technologies, microwave has become a new solution for bearer networks. Generally, microwave transmission has lower latency and lower OPEX than optical transmission \cite{Parvez2018A}.
\subsubsection{Transmission technologies}
The transmission technologies of bearer networks have experienced the evolution of plesiochronous digital hierarchy (PDH), synchronous digital hierarchy (SDH), multi-service transmission platform (MSTP), packet transport network (PTN), optical transport network (OTN) and passive optical network (PON). The main purpose of these technological evolution is not to lower latency (it pay more attention to capacity, bandwidth and cost), but it still have some impact on latency. The impact of these transmission technologies on latency in the evolution process can be summarized as follows:
\begin{itemize}
  \item From PDH to SDH, the rate standard is standardized, the interface is unified, and the management capability is enhanced.
  \item From SDH to MSTP, the ip-based interfaces (IP over SDH) are implemented to enhance the capacity of multi-service bearing and scheduling.
  \item From MSTP to PTN, by adopting multi-protocol label switching (MPLS) \footnote{MPLS establishes a label forwarding channel (label switching path, LSP) for messages through pre-assigned labels. At each device in the channel, only quick label switching is required (one lookup), thus saving processing time.}, the processing delay of forwarding between devices is reduced.
  \item From PTN to OTN, integrating SDH and wavelength division multiplexing (WDM) technologies, OTN realizes optical crossover instead of fiber hopping to provide more flexible scheduling while providing large capacity for long-distance transmission. By enhancing packet processing and routing forwarding capabilities, OTN can meet the needs of 5G bearer network, such as large bandwidth, low latency, high reliability, network slicing and so on.
  \item From OTN to PON, PON replaces electrical devices with optical devices in order to realize all-optical network, and thus reduces the latency of electrical/optical and optical/electrical conversion between devices.
\end{itemize}

In addition to the above mainstream technologies, 5G has also emerged new bearer network technologies to reduce latency as follows:
\begin{itemize}
  \item Cut-through switching: The traditional data forwarding method is that the port checks and forwards after obtaining a complete data packet, which will introduce partial latency. Cut-through switching is the fastest forwarding mode for a switch. After receiving the destination MAC address of a data frame, the switch immediately forwards data to the destination port. Subsequent data is forwarded one byte at a time, which greatly reduces the serial forwarding latency.
  \item Flexible ethernet (FlexE): FlexE achieves physical isolation between sub-MACs and guarantees low-latency service bandwidth. At the same time, the low-latency identifications of FlexE can be passed to network processor (NP), traffic manager (TM) so as to achieve E2E low-latency channel.
  \item Latency-aware priority: Optimizing NP kernel to sense the priority of low-latency services, a dedicated channel for low-latency services can be opened up.
  \item Priority-based latency scheduling: According to the priority of the delay traffic from NP, TM adopts message-through scheduling and preemptive scheduling mechanism to guarantee the requirement of low-latency services.
\end{itemize}

\section{Evolutionary Physical Layer Solutions for Low Latency}\label{Sec:PHY}
In order to achieve low latency, it is not only necessary to change the network architecture, but also the wireless air interface technologies. In this section, we mainly focus on the evolution of physical layer technologies to reduce $T_{PL}$, including frame structure, scheduling, multiple access, modulation, channel coding and signal carrier. The main evolutionary physical layer solutions for low latency are summarized in \textbf{Table \ref{table:PHY}}.
%
%
%
%

\newcommand{\tabincell}[2]{\begin{tabular}{@{}#1@{}}#2\end{tabular}}
\begin{table*}[htbp]
\centering
\caption{Summary of evolutionary physical layer technologies for low latency.}
\label{table:PHY}
\begin{tabular}{c|c|l|c|c|c|c|c|c|c}
\hline
\hline
\multicolumn{2}{c|}{}&\multicolumn{3}{c|}{\tabincell{c}{Frame \\structure}}&\multirowcell{2}{\tabincell{c}{ Scheduling \\schemes}} &\multirowcell{2}{\tabincell{c}{Channel \\coding}}&\multirowcell{2}{\tabincell{c}{Multiple\\ access}}&\multirowcell{2}{\tabincell{c}{Modulation}}&\multirowcell{2}{\tabincell{c}{Typical \\frequency bands}}\\
\cline{3-5}
\multicolumn{2}{c|}{}&\tabincell{c}{Frame length}&\tabincell{l}{The minimum \\scheduling unit}&\tabincell{l}{TTI}&&&&&\\
\cline{1-10}
\multirowcell{2}{\tabincell{c}{2G}}&\multicolumn{1}{c|}{GPRS}&\multirowcell{2}{\tabincell{l}{A TDMA \\frame: \\60 ms}}&\multirowcell{2}{\tabincell{l}{A RLC block}}&20 ms&\multirowcell{2}{\tabincell{l}{NAN}}&\multirowcell{2}{\tabincell{c}{Convolutional \\code}}&\multirowcell{2}{\tabincell{l}{TDMA;\\FDMA}}&\multirowcell{2}{\tabincell{l}{GMSK;\\QPSK}}& \tabincell{c}{Uplink (UL): \\890-915 MHz;}\\
\cline{2-2} \cline{5-5}  
&\multicolumn{1}{c|}{E-EDGE}&&\tabincell{l}{}&10 ms& &&&&\tabincell{c}{Downlink (DL): \\935-960 MHz}\\
\cline{1-10}
\multirowcell{3}{\tabincell{c}{\\ \\ \\ \\ \\ \\ 3G}}&\multicolumn{1}{c|}{WCDMA}&\tabincell{l}{A superframe:\\ 720 ms;\\A radio frame: \\10 ms;\\ A short frame: \\2 ms}&\tabincell{l}{R99: \\
a radio frame;\\
HSDPA:\\
a short frame}&\tabincell{l}{R99: \\
10 ms;\\
HSDPA:\\
2 ms}&\multirowcell{3}{\tabincell{l}{\\\\\\Transferring \\some radio\\
interface \\control \\functions\\ from RNC \\to base \\station}}&\multirowcell{4}{}&\multirowcell{3}{}&&\tabincell{l}{UL:\\1940-1955 MHz;\\DL: \\2130-2145 MHz}\\
\cline{2-5}\cline{10-10}
&\multicolumn{1}{c|}{TD-SCDMA}&\tabincell{l}{A superframe:\\ 720 ms;\\A radio frame: \\10 ms;\\ A subframe: \\5 ms}&\tabincell{l}{A subframe}&5 ms& &\tabincell{c}{Turbo \\code}&CDMA&\tabincell{l}{GMSK;\\QPSK;\\16 QAM}&\tabincell{l}{UL:\\1880-1900 MHz;\\DL: \\2010-2025 MHz}\\
\cline{2-5}\cline{10-10}
&\multicolumn{1}{c|}{CDMA2000}&\tabincell{l}{A superframe: \\720 ms;\\A radio frame: \\10 ms;\\A slot: \\1.67 ms;}&\tabincell{l}{A slot}&1.67 ms& &&&&\tabincell{l}{UL:\\1920-1935 MHz;\\DL: \\2110-2125 MHz}\\
\cline{1-6}\cline{8-10}
\multicolumn{2}{c|}{4G}&\tabincell{l}{A radio frame: \\10 ms;\\A subframe: \\1 ms\\$\bigtriangleup f=15 kHz$}&\tabincell{l}{A subframe}&1 ms&\tabincell{l}{Pre-scheduling;\\Semi-static\\ scheduling}&&\tabincell{l}{OFDM}&\tabincell{l}{GMSK;\\QPSK;\\16 QAM;\\64 QAM}&\tabincell{l}{UL:\\2500-2570 MHz;\\DL: \\2620-2690 MHz}\\
\cline{1-10}
\multicolumn{2}{c|}{5G}&\tabincell{l}{A radio frame: \\10 ms;\\A subframe: \\1 ms;\\The specific\\ intra-frame \\structure \\is shown in \\the \textbf{Table \ref{table:Num}}.\\}&\tabincell{c}{An OFDM\\ symbol\\(including CP)}&\tabincell{c}{See \\ \textbf{Table \ref{table:Num}}}&\tabincell{l}{Allocation\\ by group;\\Priority \\preemption \\scheduling}&\tabincell{l}{Data: \\LDPC;\\Control:\\ Polar}&\tabincell{l}{NOMA;\\IDMA;\\SCMA;\\FBMC;\\UFMC;\\GFDM}&\tabincell{l}{GMSK;\\QPSK;\\16 QAM;\\64 QAM;\\256 QAM}&\tabincell{l}{FR1: \\0.45-6 GHz;\\FR2: \\24.25-52.6 GHz }\\
\hline
\hline
\end{tabular}
\end{table*}

\subsection{Frame Structure}
In GPRS, the structure of 26-multiframe and 52-multiframe for CS in GSM is replaced by a new 52 time division multiple access (TDMA) frame structure in PS domain. This new frame structure transmits data in Radio Link Control (RLC) block mode. One RLC block contains 4 TDMA frame, and one TDMA frame contains eight time slots. All the 52 TDMA frames constitute 12 RLC blocks and 4 idle blocks. Since the duration of each slot is 0.577 milliseconds, the duration of all 52 TDMA frames is $52*0.577*8=240 ms$. Therefore, each RLC block period is $\frac{240}{12}=20 ms$. This RLC block period is referred to as the transmission time interval (TTI) in UMTS.

\begin{table}[htbp]
\centering
\caption{The supported transmission numerologies in 5G.}
\label{table:Num}
\begin{tabular}{c|c|l|c|c|c}
\hline
\hline
\tabincell{c}{ Parameter / Numerology (n)}&0&1&2&3&4\\
\cline{1-6}
\tabincell{c}{ $\bigtriangleup f$ (KHz)}&15&30&60&120&240\\
\cline{1-6}
\tabincell{c}{ A slot ($\mu$s)}&1000&500&250&125&62.5\\
\cline{1-6}
\tabincell{c}{The number of OFDM \\ symbols per slot \\(Normal CP)}&14&14&14&14&14\\
\cline{1-6}
\tabincell{c}{The effective length of\\ an OFDM symbol ($\mu$s)}&66.67&33.33&16.67&8.33&4.17\\
\cline{1-6}
\tabincell{c}{Length of a CP ($\mu$s)}&4.69&2.34&1.17&0.57&0.29\\
\cline{1-6}
\tabincell{c}{TTI ($\mu$s): The length of \\an OFDM symbol\\ (including CP)}&71.35&35.68&17.84&8.92&4.46\\
\hline
\hline
\end{tabular}
\end{table}

TTI represents the minimum data transmission time, referring to the length of an independent decoded transmission in a wireless link, and is the basic unit of resource scheduling and management. Reducing TTI is equivalent to reducing $T_{ttt}$, $T_{proc}$ and $T_{prop}$. At the same time, a shorter TTI can increase the number of physical layer retransmissions in a given time, thus ensuring link efficiency, i.e. reducing $T_{retr}$. TTI is the main source of data exchange latency, so most of the evolutionary schemes for reducing the latency in physical layer begin with reducing TTI.

Enhanced data rate for GSM evolution (EDGE) is a direct evolution of GPRS, often referred to as 2.75G \footnote{Since EDGE uses the same architecture of GPRS, we did not introduce it in the evolution of network architecture.}. In the evolutionary version of EDEG (called as E-EDGE), the original RLC block with 4 consecutive TDMA frames on a single channel is changed to 2 consecutive TDMA frames on a dual channel. As a result, the TTI is reduced from 20 ms to 10 ms.

There are many formats for 3G networks, the mainstream of which are wideband code division multiple access (WCDMA), time division - synchronous code division multiple access (TD-SCDMA), and code division multiple access 2000 (CDMA2000). The frame structure of WCDMA and TD-SCDMA is composed of superframe and radio frame. A superframe consists of 72 radio frames. Each radio frame lasts for 10 ms. In the 3rd Generation Partnership Project (3GPP) Release99 version of WCDMA, the minimum unit of resource scheduling is frame length, i.e., TTI is 10ms. And in the 3GPP Release5 version, the high speed downlink packet access (HSDPA) is applied to WCDMA, which introduces a short frame structure. The duration of each short frame is 2 ms, which is the minimum unit of resource scheduling, that is, TTI is reduced to 2 ms. TD-SCDMA introduces subframes as resource scheduling units, each of which has a length of 5 ms, i.e., TTI is 5 ms. In CDMA 2000, the radio frame is composed of 16 slots. The slot length of 1.67ms is the basic unit of resource scheduling, so TTI is 1.67ms. In summary, the minimum TTI that 3G can achieve is 1.67 ms.

In Long Term Evolution (LTE) systems, the radio frame structure is also adopted. The radio frame length is 10 ms and consists of two half-frames with a length of 5ms. Each half-frame consists of five sub-frames with a length of 1 ms, including four ordinary sub-frames and one special sub-frame. Therefore, the whole frame can also be understood to be divided into 10 sub-frames with length of 1ms as the unit of data scheduling and transmission (i.e., TTI).

Following previous generations of mobile communications, 5G should adopt a shorter sub-frame structure. But unlike many people's expectations, 5G still adopts the same 1ms sub-frame as 4G LTE . This is mainly due to the long-term existence of LTE, so 5G needs to consider the compatibility of new radio (NR) and LTE. In order to achieve low latency, the compromise is that the number of orthogonal frequency division multiplexing (OFDM) symbols in a sub-frame is no longer always 14. In 5G NR, resources are scheduled in units of OFDM symbols instead of sub-frames. That is, the length of TTI depends on the length and number of OFDM symbols. 5G NR support multiple numerologies (including subcarrier spacing and symbol length) for different services. There is only one slot fixed in each subframe in LTE, whereas the number of slots contained in NR subframe is related to the specific numerology. LTE uses a fixed subcarrier spacing of 15KHz, while the subcarrier spacing in NR is $\bigtriangleup f=15*2^{n}$ KHz, $n\in \{0, 1, 2, 3, 4\}$. The OFDM symbol duration is $\frac{1}{15*2^{n}} ms, n\in \{0, 1, 2, 3, 4\}$. Therefore, TTI consists of $m$ OFDM symbol duration and the duration of cyclic prefix (CP). The specific parameter sets (i.e., numerologies) are shown in the \textbf{Table \ref{table:Num}}.

In addition, 5G NR uses a more efficient mechanism to achieve low latency, that is, the so-called ``mini-slot" transmission mechanism. This ``mini-slot" mechanism allows one part of a slot to be transmitted at a time. A mini-slot even has only one OFDM symbol. This transmission mechanism can also be used to change the order of data transmission queues, so that the "mini-slot" transmission data is immediately inserted in front of the existing conventional slot transmission data sent to a terminal, so as to obtain very low latency.

\subsection{Scheduling Schemes}
Resource scheduling latency is also an important component of air interface latency, and a fast scheduling scheme can greatly reduce $T_{que}$, $T_{proc}$ and $T_{retr}$. Fast scheduling is first proposed in the 3G HSDPA, which achieves more efficient scheduling and faster retransmit by transferring some radio interface control functions from RNC to base station (closer to air interface and shorter frame length make base station scheduling faster and more efficient).

In 4G LTE before Release 14, equipment manufacturers generally used pre-scheduling to improve latency. The main idea of this method is that base stations periodically allocate corresponding wireless resources to terminals, and the terminals can send data directly on pre-allocated wireless resources when they have data to send. No need to request resources from the network side, so it reduces the time of the whole resource request process. But this method has some disadvantages:
Whether or not users use pre-scheduled wireless resources, they are always allocated to terminals. After receiving the wireless resource scheduling, if there is no data to transmit, the terminal will always upload the padding data using the allocated wireless resources. This results in the waste of precious wireless resources, the power consumption of equipments and the increase of noise level.

In view of this, in 2016, 3GPP proposed semi-static scheduling in Release 14 to improve pre-scheduling. In semi-static scheduling, even if terminals are allocated wireless resources, they do not need to send padding data.

In 4G, semi-static scheduling resources are generally allocated to each user individually. Therefore, when there are many users in the network, the waste will be very large, because the terminal does not necessarily use the reserved wireless resources. In 5G, reserved resources can be allocated to a group of users, and a collision resolution mechanism is designed when multiple users collide on the same wireless resources at the same time. In this way, the utilization of precious wireless resources is guaranteed while reducing the latency.
In addition, the priority preemption scheduling can be used into 5G networks to ensure the latency requirement of low-latency services. If idle resources in time and frequency domain are available, the user B with high priority will be given priority in scheduling idle resources. Without idle resources available, user B will preempt the resources of other users (e.g. user A), even if the user A has been originally scheduled in the corresponding slot.

\subsection{Channel Coding Schemes}
In 1949, R. Hamming and M. Golay proposed the first practical error control coding scheme, i.e., Hamming code. Subsequently, with the help of cyclic shift, cyclic redundancy check (CRC) code was proposed, which greatly reduces the coding and decoding structure and reduces the coding and decoding latency.

However, Hamming and CRC coding schemes are both based on block codes. There are two main shortcomings of block codes: one is that the decoding process must wait for the whole codeword to receive before it can begin to decode; the other is that accurate frame synchronization is needed, which leads to large latency and large gain loss.

In 1955, Elias proposed convolutional coding that makes full use of the correlation among various information blocks. In the decoding process of convolutional codes, not only the decoding information is extracted from the code, but also the relevant information of decoding is extracted from the codes received before and after. And the decoding is carried out continuously, which can ensure that the decoding latency of convolutional codes is relatively low. Convolutional codes also have the problem of computational complexity, and there is always a gap of 2-3dB between their gain and Shannon's theoretical limit.

Combining convolutional codes with interleavers, the parallel cascade convolutional code, namely Turbo, was proposed in 1993. By exchanging reliability information iteratively to improve its decoding results, Turbo code achieves performance close to Shannon limit. But Turbo code does not solve the problem of complexity, and its complexity increases with the increase of interleaving depth.

In the case of high real-time requirement, Turbo code encounters bottlenecks for the upcoming 5G demand of ultra-high speed and ultra-low delay. Therefore, in the 5G era, there is a dispute between Polar code and low-density parity check (LDPC) code. LDPC code was proposed by MIT professor Robert Gallager in 1962, which was the first proposed channel code approaching shannon limit. LDPC is based on efficient parallel decoding architecture and its decoder is superior to Turbo codes in terms of hardware implementation complexity and power consumption. Polar code was proposed by Professor E. Arikan of Bilken University in Turkey in 2007. It is a coding scheme that has been proved theoretically to reach Shannon limit. Polar codes have lower coding and decoding complexity, and there is no error floor phenomenon. The frame error rate (FER) is much lower than Turbo's. Polar codes also support flexible encoding lengths and rates, and have proven to be better than Turbo codes in many aspects.

Finally, 3GPP abandoned Turbo code in 5G era and chose LDPC as data channel coding scheme and Polar as broadcast and control channel coding scheme. Due to the different advantages and disadvantages of various coding schemes, the hardware implementation complexity, power consumption, flexibility and maturity should be comprehensively considered.

\subsection{Multiple Access and Modulation}
The development of modulation technology and multiple access technology is mainly to improve the efficiency of spectrum utilization, that is, to increase the amount of data transmitted under the same spectrum bandwidth per unit time. In the case of large data and users, high spectrum utilization efficiency can greatly reduce queuing latency $T_{que}$, and then reduce the overall latency. It is noteworthy that high-order modulation will inevitably lead to high complexity and thus increase processing latency. Therefore, to reduce the delay, the choice of appropriate modulation mode needs to weigh $T_{que}$ and $T_{proc}$ according to application scenarios.

In addition, some new low-latency multiple access technologies have emerged in 5G, such as interleave division multiple access (IDMA), sparse code multiple access (SCMA), non orthogonal multiple access (NOMA), filter bank multi carrier (FBMC), universal filtered multi-carrier (UFMC) and generalized frequency division multiplexing (GFDM). These multiple access technologies have the following characteristics in reducing latency:
\begin{itemize}
  \item IDMA simplifies the complexity of multi-user detection (MUD) without complex transmission scheduling strategy, and thus reduce the $T_{proc}$ and $T_{que}$.
  \item Combining symbol mapping and spreading and introducing non-orthogonal sparse code domain, SCMA achieves three times the number of connections. At the same time, because SCMA allows users to have certain conflicts, the application of the dispatch-free technology in SCMA can significantly reduce data transmission latency.
  \item The synchronization requirement of NOMA receiving algorithm for different signal arrival time is not high, which makes terminals can send data directly without waiting for the base station to allocate dedicated uplink resources. Compared with traditional scheduling-based resource allocation, the NOMA technology can save a request scheduling and scheduling authorization cycle, save time and network resources.
  \item FBMC, UFMC and GFDM are all based on filters, which can all improve spectral efficiency and reduce the latency mainly by shortening CP and reducing the dependence of synchronization. The long transmission impulse response length leads to the long frame length of FBMC, although the CP is shortened. In addition, the computational complexity of FBMC is much higher than OFDM, which makes it unsuitable for low-latency communication. Thus, UFMC has improved FBMC by filtering through a set of continuous subcarriers. GFDM replaces linear convolution with cyclic convolution, which reduces computational complexity and processing latency.
  \end{itemize}

\subsection{Signal Carrier}
With the development of mobile communication technology, carrier frequencies are increasing (from 800-900 MHz of 2G to  millimeter-wave bands of 5G). With the increase of frequency, the number of sinusoidal waves per unit bandwidth will increase, i.e., the amount of information carried by unit bandwidth will increase. Therefore, $T_{que}$ can be reduced when the number of data and users is large.

In addition, the coverage of base stations becomes smaller at the same power due to the increase of frequency. As a result, the cell size is shrinking, which reduces the propagation latency $T_{prop}$.

\begin{table*}[htbp]
\centering
\caption{Summary of low latency evolution schemes.}
\label{table:SLLES}
\begin{tabular}{c|l|l|l|l}
\hline
\hline
\multirow{2}{*}{}&\multicolumn{3}{c|}{Network architecture}&\multirowcell{2}{\tabincell{c}{Physical layer technologies}}\\
\cline{2-4}
&\multicolumn{1}{c|}{RAN}&\multicolumn{1}{c|}{Core network}&\multicolumn{1}{c|}{Bear network}&\\
\cline{1-5}

3G&\tabincell{l}{1. Replacing private interface Abis\\~~ with open interface Iub;\\2. Sinking several control functions\\~~ from RNC to NodeB;\\3. Removing the PCU entity;\\ 4. The interface A and Gb are \\~~~consolidated into the interface Iu;\\ 5. The interface Iur between RNCs \\~~~is added;\\ 6. Some radio-related management \\~~~functions are moved from core \\~~~network into RAN;\\7. All interfaces realize the separation \\~~~of control plane and user plane at \\~~~protocol level.}
&\tabincell{l}{1. The complete separation \\~~~of bearer and control \\~~~in CS domain, logically \\~~~and physically.  }
&\multirowcell{3}{\tabincell{l}{1. The transmission \\~~~distance is continuously \\~~~shortening;\\2. The transmission medium \\~~~has evolved from \\~~~copper wire to optical \\~~~cable and then to \\~~~microwave transmission;\\3.  The transmission technologies \\~~~have experienced the \\~~~evolution of
PDH, SDH, \\~~~MSTP, PTN, OTN and PON; \\4. All-IP; \\5. All-optical;\\6. Cut-through switching;\\7. FlexE;\\8. Latency-aware priority;\\9. Priority-based latency \\~~~scheduling.}}

&\multirowcell{3}{\tabincell{l}{1. Shorter TTI;\\2. Faster scheduling;\\3. Coding with lower complexity \\~~~and higher efficiency; \\4. Multiple access technology \\~~~with higher spectrum \\~~~utilization efficiency; \\5. Higher-order modulation;\\6. Higher frequency carrier.}}\\

\cline{1-3}
4G&\tabincell{l}{1. Removing RNCs;\\2. The interface X2 between \\~~~eNodeBs is added;\\3. Sinking several control functions \\~~~from RNC and \\~~~core networks to eNodeB;\\ 4. The concept of C-RAN \\~~~was introduced to improve\\~~~resource scheduling capability.}
&\tabincell{l}{1. Removing the CS domain; \\2. The control plane and user \\~~~plane are completely \\~~~separated, physically and \\~~~logically.}
&
&\\

\cline{1-3}
5G&\tabincell{l}{1. The network elements is able \\~~~to move more flexibly;\\2. The reconstructed three \\~~~functional entities promote \\~~ RAN network slicing.;\\3. NFV and SDN build VCRPs \\~~~to speed up network sessions;\\ 4. Introducing fog computing \\~~~into RAN to build F-RAN;\\ 5. D2D communication directly.}
&\tabincell{l}{1. Traditional network entities \\~~~are split into multiple \\~~ virtualized NFs by NFV \\~~~and SDN; \\2. The traditional point-to- \\~~~point communication \\~~ between
network \\~~~elements is replaced by \\~~~bus communication mode;\\3. The user plane functions \\~~~can further sink to MEC \\~~~nodes. }
&&\\

\hline
\hline
\end{tabular}
\end{table*}

\section{Evolutionary Medium Access Control Layer for Low Latency}\label{Sec:MAC-layer}
The medium access control (MAC) layer is responsible for resource scheduling, multiple
access, mobility management, interference management, rate adaptation, and synchronization. A good design of MAC layer technologies can greatly reduce network latency. In this section, we review the evolution of resource scheduling schemes, multiple access technologies, mobility management and caching technologies to help reduce the latency.
\subsection{Scheduling Schemes}
Resource scheduling latency is also an important component of air interface latency, and a fast scheduling scheme can greatly reduce $T_{que}$, $T_{proc}$ and $T_{retr}$. Fast scheduling is first proposed in the 3G HSDPA, which achieves more efficient scheduling and faster retransmit by transferring some radio interface control functions from RNC to base station (closer to air interface and shorter frame length make base station scheduling faster and more efficient).

In 4G LTE before Release 14, equipment manufacturers generally used pre-scheduling to improve latency. The main idea of this method is that base stations periodically allocate corresponding wireless resources to terminals, and the terminals can send data directly on pre-allocated wireless resources when they have data to send. No need to request resources from the network side, so it reduces the time of the whole resource request process. But this method has some disadvantages:
Whether or not users use pre-scheduled wireless resources, they are always allocated to terminals. After receiving the wireless resource scheduling, if there is no data to transmit, the terminal will always upload the padding data using the allocated wireless resources. This results in the waste of precious wireless resources, the power consumption of equipments and the increase of noise level.

In view of this, in 2016, 3GPP proposed semi-static scheduling in Release 14 to improve pre-scheduling. In semi-static scheduling, even if terminals are allocated wireless resources, they do not need to send padding data.

In 4G, semi-static scheduling resources are generally allocated to each user individually. Therefore, when there are many users in the network, the waste will be very large, because the terminal does not necessarily use the reserved wireless resources. In 5G, reserved resources can be allocated to a group of users, and a collision resolution mechanism is designed when multiple users collide on the same wireless resources at the same time. In this way, the utilization of precious wireless resources is guaranteed while reducing the latency.
In addition, the priority preemption scheduling can be used into 5G networks to ensure the latency requirement of low-latency services. If idle resources in time and frequency domain are available, the user B with high priority will be given priority in scheduling idle resources. Without idle resources available, user B will preempt the resources of other users (e.g. user A), even if the user A has been originally scheduled in the corresponding slot.

\subsection{Multiple Access and Modulation}

 The purpose of multiple access technologies is to enable multiple users to access the base station at the same time and enjoy the communication services provided by the base station, so as to ensure that the signals between each user will not interfere with each other. Each generation of communication system has its own unique multiple access technology. The techniques can be divided into two categories \cite{tootoonchian2012controller}: contention-free MAC protocols and contention-based MAC protocols.

\subsubsection{Contention-free MAC Protocols}
 The FDMA, TDMA, CDMA and OFDM are employed by 1G, 2G, 3G and 4G, respectively. The original intention of these multiple access technologies is not to reduce the latency, but to improve the system capacity and accommodate more users. In terms of latency, FDMA has more advantages, because of its low cost of continuous transmission, no need of complex framing and synchronization, no need of channel equalization and so on. On the contrary, TDMA, CDMA and OFDM have more advantages than FDMA in system capacity and access.

In the age of 5G, latency becomes an important consideration. Thus, some new low-latency multiple access technologies have emerged in 5G, such as interleave division multiple access (IDMA), sparse code multiple access (SCMA), non-orthogonal multiple access (NOMA), filter bank multi carrier (FBMC), universal filtered multi-carrier (UFMC) and generalized frequency division multiplexing (GFDM). These multiple access technologies have the following characteristics in reducing latency:
\begin{itemize}
  \item IDMA simplifies the complexity of multi-user detection (MUD) without complex transmission scheduling strategy, and thus reduce the $T_{proc}$ and $T_{que}$.
  \item Combining symbol mapping and spreading and introducing non-orthogonal sparse code domain, SCMA achieves three times the number of connections. At the same time, because SCMA allows users to have certain conflicts, the application of the dispatch-free technology in SCMA can significantly reduce data transmission latency.
  \item The synchronization requirement of NOMA receiving algorithm for different signal arrival time is not high, which makes terminals can send data directly without waiting for the base station to allocate dedicated uplink resources. Compared with traditional scheduling-based resource allocation, the NOMA technology can save a request scheduling and scheduling authorization cycle, save time and network resources.
  \item FBMC, UFMC and GFDM are all based on filters, which can all improve spectral efficiency and reduce the latency mainly by shortening CP and reducing the dependence of synchronization. The long transmission impulse response length leads to the long frame length of FBMC, although the CP is shortened. In addition, the computational complexity of FBMC is much higher than OFDM, which makes it unsuitable for low-latency communication. Thus, UFMC has improved FBMC by filtering through a set of continuous subcarriers. GFDM replaces linear convolution with cyclic convolution, which reduces computational complexity and processing latency.
  \end{itemize}


\section{Open Issues of Network Architecture and Physical Layer for Low Latency}\label{Sec:Openissues}
Although there are many schemes to reduce latency, there are still some research directions and challenges waiting for researchers to explore and solve. In this section, we discuss some open issues and challenges
for future research to further reduce latency.
\subsection{Network Architecture Issues}
To achieve low latency, it is often necessary to make tremendous changes to the existing network architecture. In the process of network reconfiguration, many challenges need to be overcome, which can be summarized as follows:
\begin{itemize}
  \item For communication networks, practicality is the key. In order to truly apply a network architecture that can reduce the latency to real life, the cost problem must be considered. How to achieve low latency through the change and deployment of network architecture at low cost is worth studying. One of the best ways is to take advantage of the existing architecture.
  \item From 2G to 5G, the current mobile communication network is a heterogeneous network constructed by a variety of network systems. How to manage heterogeneous networks in order to achieve efficient utilization of resources while maintaining low latency is also a research area.
  \item From the previous description, we can conclude that the network is becoming software and virtualization. These new virtual entities implemented by SDN and NFV technology are totally different from legacy networks, and they have not yet unified standards. Extensive research needs to be studied on how to standardize and unify them.
  \item Unmanned aerial vehicles (UAVs) and satellites can be integrated into traditional cellular networks to reduce latency. However, resource management and interference control need to be addressed.
  \item Future networks will become ultra-dense cellular networks with numerous small cells, which makes it possible for millimeter-wave wireless bearers. The challenge is to design new bearer networks and protocols to overcome interference and collision. In addition, cooperative co-transmission between small cells can also be a research direction to reduce latency. There are many issues worth studying about this direction, such as how to achieve dynamic cooperation and how to reduce overhead caused by inter-cell interaction.
\end{itemize}
\subsection{Physical Layer Technique Issues}
Most of physical layer technologies are not designed to reduce the latency. Thus, these original physical layer technologies must be redesigned to reduce the latency. Although many promising solutions have been proposed to date, we believe that the following issues about low latency at physical layer that deserve further exploration:
\begin{itemize}
  \item Reducing latency may cause other performance degradation. For example, low latency is related to control overhead (including CP, pilot, etc). Short TTI means that the control overhead proportion increases, resulting in the waste of radio frequency resources. Therefore, various trade-offs need to be investigated, including spectrum efficiency versus latency, energy efficiency versus latency, and throughput versus latency.
  \item Fast channel estimation algorithm and efficient symbol detection are conducive to reducing latency. However, with the emergence of multi-antenna, high-order modulation and new multiple access technologies, these algorithms for scheduling also need to be redesigned to reduce latency.
  \item With the advent of 5G, almost no single physical layer technology can meet the requirements of all service scenarios. How to dynamically, efficiently and intelligently coordinate various physical layer technologies to meet the needs of different service scenarios is a potential research direction. Therefore, the application of machine learning in the physical layer has also attracted wide attention.
  \item Millimeter wave is one of the promising technologies of 5G, but many of its channel characteristics have not been fully understood. Deep understanding of the characteristics of attenuation, angular spread, reflection, Doppler effect and atmospheric absorption is helpful to the design of appropriate physical layer technologies.
  \item The propagation characteristics of short packet transmission (with shorter TTI) and traditional packet transmission in channel are different. In the case of large packet transmission, distortion and thermal noise can be averaged. In the case of large packet transmission, thermal noise can be averaged, but small packet transmission is not feasible. Therefore, channel modeling and experiment under short packet transmission need to be further explored.
\end{itemize}

\section{Conclusion}\label{Sec:Conclusion}
In this paper, we review the measures to reduce the latency of 2G to 5G cellular network communication from an evolutionary perspective, including network architecture and physical layer technologies. In order to achieve low latency, on the one hand, we can draw the following conclusions in the evolution of network architecture: 1) fewer network units, that is, the Architecture tends to be flat; 2) network elements are gradually transformed from entity to virtual unit, that is virtualization; 3) data and signaling transmission control is gradually transformed from hardware to software, that is, software; 4) functional entities continue to sink closer to terminals; 5) network bearer gradually unified (all-optical and all-IP).  On the other hand, we can draw the following conclusions from the evolution of physical layer technologies: 1) TTI is decreasing continuously to achieve low latency in a smaller packet transmission mode; 2) scheduling algorithm is becoming faster and faster, even dispatch-free transmission; 3) coding mode is more flexible and changeable, no longer one-size-fits-all; 4) Multiple access technologies and modulation methods tend to have more latitudes and higher orders to achieve huge connections and thus reduce queue latency; 5) Carrier frequency is higher and higher, and cell size is smaller and smaller.
Although we can see many solutions to reduce latency through our review, there are still many difficulties and challenges to be solved in practical application. Thus, we also give some challenges and future research topics, hoping this can be served to reduce latency for the next generation mobile communication system. The low latency evolution schemes involved in this paper are summarized in \textbf{Table \ref{table:SLLES}}.

\section*{Acknowledgments}
This work was partly supported by the National Natural Science Foundation of China (Grant Nos. 61871023 and 61931001), Beijing Natural Science Foundation (Grant No. 4202054), and the Fundamental Research Funds for the Central Universities (Grant No. 2019YJS010).

\bibliographystyle{IEEEtran}
\bibliography{ref}
\end{document}